\documentclass{article}

\usepackage{amsfonts}
\usepackage{float}
\usepackage{amsmath}
\usepackage{amssymb}
\usepackage{amsthm}
\usepackage{graphicx}
\usepackage{multirow}
\usepackage{color}
\usepackage{ulem}
\usepackage{setspace}
\usepackage{caption}
\usepackage{subcaption}
\usepackage{rotating}
\usepackage{threeparttable}
\usepackage{url}
\usepackage{framed}
\usepackage[round]{natbib}

\newcommand{\mbf}[1]{\mbox{\boldmath $#1$}}

\def\bA{\mathbf{A}}

\def\bP{\mathbf{P}}
\def\bR{\mathbf{R}}

\def\bu{\mbf{u}}

\def\bW{\mathbf{W}}
\def\bX{\mathbf{X}}
\def\bG{\mathbf{G}}

\def\bD{{\mathbf D}}

{\catcode `\@=11 \global\let\AddToReset=\@addtoreset}
\AddToReset{equation}{section}

\AddToReset{Theorem}{section}

\newtheorem{lem}{Lemma}[section]
\newtheorem{thm}{Theorem}[section]

\theoremstyle{remark}
\newtheorem{rem}{Remark}[section]

\newcommand{\cB}{{\cal B}}

\newcommand{\cD}{{\cal D}}

\newcommand{\cF}{{\cal F}}

\newcommand{\cI}{{\cal I}}

\newcommand{\cL}{{\cal L}}
\newcommand{\cM}{{\cal M}}
\newcommand{\cN}{{\cal N}}

\newcommand{\cQ}{{\cal Q}}

\newcommand{\cW}{{\cal W}}

\def\bc{\begin{center}}
\def\bd{\begin{description}}
\def\be{\begin{enumerate}}
\def\ec{\end{center}}
\def\ed{\end{description}}

\def\ee{\end{enumerate}}
\def\ben{\begin{equation}}
\def\benn{\begin{equation*}}
\def\een{\end{equation}}
\def\eenn{\end{equation*}}
\def\benr{\begin{eqnarray}}
\def\eenr{\end{eqnarray}}
\def\benrr{\begin{eqnarray*}}
\def\eenrr{\end{eqnarray*}}

\def\b{\beta}

\def\D{\Delta}

\def\edt{\end{document}}

\def\G{\boldsymbol{\Gamma}}

\def\iny{\infty}

\def\la{\lambda}

\def\lel{\label}

\def\noi{\noindent}

\def\ra{\rightarrow}

\def\si{\sigma}

\def\sti{\sum_{i=1}^n}

\def\stj{\sum_{j=1}^p}

\def\stk{\sum_{k=0}^\iny}

\def\vep{\varepsilon}

\def\vp{\varphi}

\def\wh{\widehat}

\def\wt{\widetilde}

\def\bx{\mbf{x}}

\def\bxi{\mbf{x}_{i}}
\def\bxii{\mbf{x}_{1i}}
\def\bxiii{\mbf{x}_{2i}}

\def\bSn{\mbf{S}_{n}}
\def\bqi{\mbf{q}_{i}}

\def\bu{\mbf{u}}

\def\bg{\mbf{g}}

\def\bth{\mbf{\theta}}
\def\bTh{\mbf{\Theta}}

\def\B{\boldsymbol{\beta}}
\def\D{\boldsymbol{\delta}}

\def\W{\boldsymbol{\cW}}
\def\S{\boldsymbol{\Sigma}}

\def\bL{\boldsymbol{\Lambda}}
\def\bO{\boldsymbol{\Omega}}

\def\bGa{\boldsymbol{\Gamma}}

\def\mR{\mathbb{R}}
\def\mP{\mathbb{P}}
\def\mE{\mathbb{E}}

\def\tbf{\textbf}

\newcounter{Qcounter}
\newcounter{show}
\title{Robust and Efficient Estimation for Count Data Using $L_{2}$ Optimization}
\author{Jiwoong Kim\\
University of South Florida}
\date{}
\begin{document}
\maketitle
\begin{abstract}
This paper proposes a novel method to estimate the parameter of Poisson-distribution-related regressions. The proposed method employs the Cramer-von Mises type optimization which has been commonly used in estimating parameters of continuous distributions. Upon obtaining the estimator through the proposed method, its desirable properties such as asymptotic distribution and robustness are rigorously investigated. Simulation studies serve to demonstrate that the proposed method compares favorably with other well-celebrated methods including the maximum likelihood method.
\end{abstract}
\noi


\section{Introduction}
The Poisson probability distribution is one of many distributions used for modelling count data such as the number of car accidents, the number of typos on a page, etc. The most prominent feature of the Poisson distribution is that its mean and variance take the identical value, which is the rate parameter $\la$; this property is referred to as \textit{equidispersion}, and, the Poisson distribution of the count data is governed solely by a single parameter, which is a great merit in terms of estimation. On the regression setup, the logarithm of the rate parameter -- which is called a link function -- is equated with either linear or non-linear function of predictors and unknown regression parameters. For example, the rate of car accidents can be influenced by driver's experience, gender, etc. In real-world application, the equidispersion is a rare assumption, as many sample count data exhibits quite large discrepancy between the sample mean and variance. The case of larger variance is referred to as overdispersion, while the opposite case is called underdispersion. \cite{Consul} proposed the generalized Poisson (GP) distribution and demonstrated that incorporating another parameter with the probability mass function can handle both cases; see Section \ref{sec:GP_distribution} for more details. Based on the GP distribution, \cite{Yang2009} proposed a new score test for overdispersion and demonstrated that it compares favorably with other well-celebrated tests.

Minimum distance (MD) estimation is a classical methodology that has been used in many statistical inference problems, including estimation and hypothesis testing. While the MD estimation has many interesting features, the most prominent one among many is its distance function, which is designed to measure deviation of data points from the assumed model. To that end, the distance function takes difference between empirical distribution function (df) and the modeled df, which represent the observed data points and assumed model, respectively; thus, as the parameter estimation gets more accurate, the difference between empirical and modeled df's gets smaller, leading to smaller quantity of the distance function, which is the quintessence of the MD method. Similarly, as the difference between two functions is enlarged, the quantity of the distance function also increase, which provides a felicitous extension of application of the MD method a hypothesis testing.

Having been popular with statisticians, the statistical literature is awash with research works on the MD method; for more details such as references for those research works, see Section \label{sec:literature_review}. However, research on the MD method had dealt with continuous cases only -- that is, one-sample parameter estimation with continuous random variables, regression parameter estimation with a continuous response variable, etc -- before \cite{Kim2026} proposed a modified MD method designed for estimating a parameter of a discrete probability distribution; the distance function used in his work still measured the discrepancy between the empirical and modeled df's but employed the sum, instead of integral, to deal with discrete random observations; it transpired that all desirable properties -- including asymptotical normality and robustness -- that the MD method exhibits when applied to the continuous cases were preserved despite the modification. Motivated by this fact, we enlarge application of the MD method from one-sample setup to regression setup of the Poisson distribution. Many results of this article have a root in \cite{Kim2026}. However, it should be noted that most works in this article is not a simple replica of his work in that there is another major modification in the distance function, and hence, the subsequent analysis thereafter; see Section \ref{Sec:MD_one_sample} for more details. We expect that the MD estimation still retains those desirable properties when it is applied to the estimation of regression parameters of both original and GP distributions.

The rest of this article is organized as follows. Starting with the literature review in Section \ref{sec:literature_review}, Section \ref{Sec:MDE} proposes a new MD estimation method based on the $L_{2}$ optimization theories; Sections \ref{Sec:MD_one_sample} and \ref{Sec:MD_regression} define the estimation problems in details for the the one-sample and regression setups of the Poisson distribution, respectively; focusing, rather, on the regression setup, Sections \ref{sec:MD_regression_estimator}, \ref{sec:MD_robustness}, and \ref{sec:MD_bias_reduction} propose the new estimator of the Poisson regression parameter and derive its asymptotic properties such as asymptotic normality, robustness, and bias reducibility. Section \ref{Sec:GP_Regression} investigates further application of the proposed method to the GP regression. For all the statistical analyses in this article, we use an \texttt{R} package \texttt{jwPois}, which is available in \url{https://github.com/jwboys26/jwPois}.

\section{MD estimation}\label{Sec:MDE}
\subsection{Selection of the distance function}\label{sec:literature_review}
Since it was introduced by \cite{Wolfowitz1953}, the MD method has been popular due to its desirable properties such as asymptotic normality and robustness. During 1970s and 1980s, colossal amount of research has been conducted on it: see \citet[Chapter 5]{Koul2002} and references therein. For these research works, the selection of the distance function played an important role to derive the desirable results. For example, \cite{Parr} demonstrated that the MD estimation using the Cramer von-Mises (CvM) $L_{2}$-distance yields the more robust estimator than those using other distance functions. \cite{Koul2002} also demonstrated that the CvM-type distance function with various integrating measures, including Lebesgue, dirac, and probability measures, yields the most efficient estimator of the regression parameter when the independent error terms of the regression model are normal, logistic, or laplace random variables, respectively. Furthermore, \cite{Kim2020} applied the MD estimation with the CvM $L_{2}$-distance function to a regression model with dependent errors and demonstrated that the original desirable properties that hold under the assumption of independent errors are still reserved. Motivated by this fact, we will also use an analogue of the CvM $L_{2}$-distance function where the integral of the original distance function is replaced by the sum for the Poisson distribution.

Since 1990, the popularity of the MD methodology waned rapidly; only a few statisticians had conducted research on it. The main culprit of rapid decline in its popularity is attributed to the complexity to compute the MD estimator. For this reason, the research on the MD estimation focused on its computational aspects, rather than on the theoretical aspects. Reflecting this trend, \cite{Dhar1991, Dhar1992} and \cite{Kim2018} studied computational aspects of the MD estimators for regression and autoregressive models where error and innovation terms are independent.

Aforementioned research dealt with continuous setup where the response variable of regression models is continuous random variable. As mentioned in the introduction, \cite{Kim2026} extended application of the MD estimation from continuous distributions to discrete distributions.

\subsection{MD estimation on an one-sample setup}\label{Sec:MD_one_sample}
Consider a random sample of independent and identically distributed (iid) Poisson random variables $X_{1},...,X_{n}$.  Let $f$ and $F$ denote the Poisson probability mass function (pmf) and distribution function (df) with a true rate parameter $\la_{0}$, respectively. Then, $f$ is expressed as
\benn
f(k;\la_{0}) = \frac{\la_{0}^{k}e^{-\la_{0}}}{k!},
\eenn
where $\la_{0}\in \mathbb{R}^{+}$ and $k\in \{0\}\cup \mathbb{N}$, while $F$ is a cumulative sum of $f$, i.e., $F(k;\la_{0})=\sum_{l=0}^{k}f(l;\la_{0})$. The problem of interest will be to estimate the unknown $\la_{0}$.

When \cite{Kim2026} defined the distance function to estimate the success probability of a binomial distribution, he used cumulative indicator and binomial df for the empirical and assumed distribution functions, respectively, in the summand. As done in \cite{Kim2026}, the distance function of $\la\in \mathbb{R}^{+}$ in this study can be defined as
\benn
L_{1}(\la) = \sum_{k=0}^{\iny}\left[ \sti d_{ni}\Big\{\textrm{I}(X_{i}\le k)-F(k;\la) \Big\} \right]^{2},
\eenn
where $\textrm{I}(\cdot)$ is an indicator function and $d_{ni}\in \mathbb{R},\,1\le i\le n$. Subsequently, the MD estimator of the rate parameter, denoted by $\wh{\la}$, can be obtained by minimizing the above distance function. At this juncture, one interesting question arises: can we use non-cumulative indicator function and pmf to define a distance function? For example, consider
\benn
L_{2}(\la) = \sum_{k=0}^{\iny}\left[ \sti d_{ni}\Big\{\textrm{I}(X_{i}= k)-f(k;\la) \Big\} \right]^{2},
\eenn
which has not been an possible option for defining the distance function for the continuous setup. 

Deriving the asymptotic properties of the MD estimator in \cite{Kim2026} required to differentiate the assumed function with respect to the parameter of interest. Since the binomial df doesn't have an analytic expression, the derivation of the asymptotic normality of the MD estimator therein became a bit complicated. Becoming the upper incomplete gamma function, the Poisson df has a similar issue; differentiating it with respect to the rate parameter will encounter an complicated integral. To address the complexity, this study will consider to embed the Poisson pmf, as well as the df, in the distance function, which renders the analysis easier. Let
\benn
\cI(X_{i},k; \cF):= \left\{
                            \begin{array}{ll}
                              \textrm{I}(X_{i}=k), & \hbox{if $\cF=f$;} \\
                              \textrm{I}(X_{i}\le k), & \hbox{if $\cF=F$.}
                            \end{array}
                          \right.
\eenn
and define $H(k, \la; \cF):=\mE[\cI(X_{1},k; \cF)]$. Note that these notations provide a convenient tool to define distance function. For example, $L_{1}$ and $L_{2}$ above can be rewritten as
\benn
L_{1}(\la) = \sum_{k=0}^{\iny}\left[ \sti d_{ni}\Big\{\cI(X_{i},k; F)-H(k,\la;F) \Big\} \right]^{2},
\eenn
while
\benn
L_{2}(\la) = \sum_{k=0}^{\iny}\left[ \sti d_{ni}\Big\{\cI(X_{i},k; f)-H(k,\la;f) \Big\} \right]^{2}.
\eenn

It turns out that the derivative of $H$ with respect to $\la$ possesses some useful properties for deriving the asymptotic normality. Let $\partial H(k,\la; \cF)$ and $\partial^2 H(k,\la; \cF)$ denote the first- and second-order partial derivatives of $H(k,\la; \cF)$ with respect to $\la$, respectively. First consider $\cF=f$: $\partial H(k,\la; f)=\partial f(k;\la)/\partial \la$ and $\partial^2 H(k,\la; f)=\partial^2 f(k;\la)/\partial \la^2$. Direct calculations show that
\ben\label{eq:gk_definition}
\partial  H(k,\la; f) = \frac{\la^{k-1}e^{-\la}}{k!}(k-\la)\,\, \textrm{ and }\,\, \partial^2 H(k,\la; f) = \partial H(k-1,\la; f)-\partial H(k,\la; f).
\een
It is worth mentioning several useful facts. First, it is easy to see that $\partial  H(k,\la; f)=f(k-1;\la)-f(k;\la)$. Second, the previous result rewrites $\partial^2 H(k,\la; f)$ as
\benn
\partial^2 H(k,\la; f) = (f(k-2;\la)-f(k-1;\la))-(f(k-1;\la)-f(k;\la)) = f(k-2;\la)-f(k;\la).
\eenn
Then,using (\ref{eq:gk_definition}), we obtain, for any bounded $\la$,
\ben\label{eq:gk_bdd}
\stk \left|\partial H(k,\la; f)\right|^{r}=O(1)\,\textrm{ and }  \stk \left|\partial^2 H(k,\la; f)\right|^{r}=O(1)\,\,\textrm{for }r=1,2.
\een
For example, when $r=2$, we have
\benrr
\stk |\partial H(k,\la; f)|^{2}&\le& 2 \stk\left(\frac{\la^{k-1}e^{-\la}}{k!}\right)^2(k^{2}+\la^{2}),\\
&=& 2 \sum_{k=1}^{\iny}\{f(k-1;\la)\}^{2} + 2\stk \{f(k;\la)\}^{2},\\
&\le & 4\stk f(k;\la)=4,
\eenrr
where $(a-b)^2\le 2(a^2+b^2)$ for $a,b\mR$ implies the first inequality while the second inequality follows from $|f(\cdot;\la)|\le 1$ and a change of variables. Using (\ref{eq:gk_definition}), The claim for the second-order partial derivative can be shown similarly. The claim of (\ref{eq:gk_bdd}) also holds for $H(\cdot, \la; F)$. To begin with, note that $\partial H(k,\la; F)=\sum_{l=0}^{k}\partial H(k,\la; f)$. Thus, for $r=2$,
\benrr
\stk |\partial H(k,\la; F)|^r&=& \stk \left|\sum_{l=0}^{k} \partial H(k,\la; f) \right|^2,\\
&=&\stk \left|\sum_{l=0}^{k} \{f(l-1;\la)-f(l;\la)\} \right|^2,\\
&=&\stk f^2(k;\la)\le 1,
\eenrr
where the second equality follows from (\ref{eq:gk_definition}).

When extending application of the MD estimation from the one-sample setup to the regression setup and deriving its asymptotic properties, we will utilize these results. As done in \cite{Kim2026}, finding the MD estimator will require some special conditions for the distance function, which will be stated in the next section.

\subsection{Extension of MD estimation to Poisson regression}\label{Sec:MD_regression}
This section will extend the application of the MD estimation to a regression setup of the Poisson distribution. Through this article, we refer to the regression setup as Poisson regression, which is a special case of a generalized linear model (GLM) used to predict count data; it models the relationship between predictors and the rate parameter through a logarithm link function. Suppose the independent count data $Y_{1},...,Y_{n}\in \{0\}\cup\mathbb{N}$ follow Poisson distributions with different rate parameters $\la_{i}$, and $Y_{i}$ are associated with predictors $\bx_{i}\in \mR^{p}$ such that,
\benn
\la_{i} = \mathbb{E}(Y_{i}|\bx_i;\B)=e^{\bx_i'\B},
\eenn
or $\log \la_{i}=\bx_i'\B$, where $\B\in \mR^{p}$ is an unknown parameter of interest. Thus, the pmf of $Y_{i}$  will be parameterized by $\bx_i,\,\B\in\mR^{p}$ and can be written as
\benn
f(k;\la_{i}) = \frac{e^{k\bx_i'\B}e^{-e^{\bx_i'\B}}}{k!},
\eenn
while its df will be expressed as 
\benn
F(k;\la_{i})=\sum_{l=0}^{k}f(k;\la_{i}).
\eenn
To stress that both $f(\cdot;\la_{i})$ and $F(\cdot;\la_{i})$ are parameterized with $\B$ and indexed by $i$ due to the independence of $Y_{i}$, we rewrite them as $f_{i}(\cdot;\B)$ and $F_{i}(\cdot;\B)$ unless they are misleading. Note that the transition from the iid condition to the independent condition of the sample causes $f$ and $F$ to be indexed by $i$.

Typical distance functions observed in the literature on MD estimation use the df, and hence, adapting those to suit the discrete setup leads to the following distance function:
\ben\label{eq:distance_function_regression}
\cL(\B)=\stj\stk\left[\sti d_{ij}\left\{ \textrm{I}(Y_{i}\le k)- F(k;e^{\bxi \B}) \right\}\right]^2,
\een
with an integral on the continuous setup being replaced by $\stk$. Note that the distance function for the current study can also be defined using $f$ as discussed in the one-sample setup in Section \ref{Sec:MD_one_sample}; in this case, the summand inside of the hard bracket will be superseded with $\textrm{I}(Y_{i}= k)- f(k;e^{\bxi \B})=\textrm{I}(Y_{i}= k)- f_{i}(k;\B)$. 

Recall $\cI$ and $H$ from the previous section. Note that \benn
\cI(Y_{i},k; \cF)= \left\{
                            \begin{array}{ll}
                              \textrm{I}(Y_{i}=k), & \hbox{if $\cF=f$;} \\
                              \textrm{I}(Y_{i}\le k), & \hbox{if $\cF=F$.}
                            \end{array}
                          \right.,\quad H(k, \la_{i}; \cF)= \left\{
                            \begin{array}{ll}
                              f_{i}(k;\B), & \hbox{if $\cF=f$;} \\
                              F_{i}(k;\B), & \hbox{if $\cF=F$.}
                            \end{array}
                          \right.
\eenn
By the same logic applied to $f_{i}$ and $F_{i}$, let $H_{i}(k, \B; \cF):=H(k, \la_{i}; \cF)$; consequently, using $\cI$ and $H_{i}$, the distance function of interest for this study can be defined as
\ben\label{eq:distance_function_regression2}
\cL(\B;\cF)=\stj\stk\left[\sti d_{ij}\left\{ \cI(Y_{i},k; \cF)- H_{i}(k,\B;\cF) \right\}\right]^2,
\een
and the corresponding MD estimator of $\B$ can be defined as
\benn
\cL(\wh{\B};\cF) = \inf_{\B\in\mathbb{R}^{p}}\cL(\B;\cF).
\eenn
Observe that the resulting MD estimator will depend on $\cF$, i.e., using a pmf for the distance function will lead to a different estimator from using a df (or vice versa).

If the most important concept in the literature on MD estimation should be chosen, that will be the the uniformly locally asymptotically quadraticity (ULAQ) of the distance function. It is not exaggeration to say that delivery of nice properties of the MD methodology, such as, asymptotic normality and robustness, will not be possible without establishing the ULAQ conditions. All analyses of this study will start from showing that the distance function in (\ref{eq:distance_function_regression}) meets the ULAQ conditions below. For the true Poisson regression parameter $\B_{0}\in \mathbb{R}^{p}$, define its neighborhood as, for $0<b<\iny$,
\ben\lel{eq:neighbood}
\cN_{b}(\B_{0}):= \{\B\in \mR^{p}:\|\bA^{-1}(\B-\B_{0})\|\le b\},
\een
where $\bA$ is some $p\times p$ symmetric, nonsingular matrix: see also \tbf{(a.1)} below. The ULAQ conditions for the Poisson regression are as follows.
\begin{itemize}
\item[(\tbf{U.1})] There exists a sequence of random vectors $\mbf{S}_{n}(\B_{0};\cF)\in \mathbb{R}^{p}$ and a sequence of $p\times p$ real matrices $\bW_{n}(\B_{0};\cF)$ such that for all $0<b<\iny$
\benn
\sup \left|\cL(\B;\cF)-\cL(\B_{0}) - 2(\B-\B_{0})'\mbf{S}_{n}(\B_{0};\cF) - (\B-\B_{0})'\bW_{n}(\B_{0};\cF)(\B-\B_{0})\right|=o_{p}(1),
\eenn
where the supremum is taken over $\B\in\cN_{b}(\B_{0})$.
\item[(\tbf{U.2})]  For all $\vep>0$, there exists a $0<c_{\vep}<\iny$ such that
    \benn
    \mathbb{P}\left( |\cL(\B_{0};\cF)|\le c_{\vep} \right)\ge 1-\vep.
    \eenn
\item[(\tbf{U.3})] For all $\vep>0$ and $0<c<\iny$, there exists a $0<b<\iny$ (depending on $c$ and $\vep$) such that
    \benn
    \liminf_{n\ra \iny}\,\mathbb{P}\left(\inf |\cL(\B;\cF)| >c \right)\ge 1-\vep,
    \eenn
    where the infimum is taken over $\{\B\in \mR^{p}:\|\bA^{-1}(\B-\B_{0})\|> b\}$.
\end{itemize}
After ascertaining that the ULAQ conditions are met, we will derive the asymptotic normality of the MD estimator. To that end, we will use Theorem 5.4.1 from \cite{Koul2002}, which is reproduced here.
\begin{lem}\label{lem:Thm541}
Suppose \textbf{(U.1)}-\textbf{(U.3)} hold. Let $\cB_{n}:=\bA\bW_{n}\bA$. Let $\wh{\B}$ denote the MD estimator that minimizes the distance function in (\ref{eq:distance_function_regression}). Then the following holds true:
\benn
\cB_{n}\bA^{-1}(\widehat{\B}-\B_{0}) = -\bA \bSn(\B_{0};\cF)+o_{p}(1).
\eenn
\end{lem}
\noi
Note that Lemma \ref{lem:Thm541} implies that deriving the asymptotic normality of the MD estimator is equivalent to deriving that of $\bA\bSn$, which will be a main task of the next section.

To prove that the distance function of this study satisfies the ULAQ conditions, we need the assumptions below. Recall $n$ pairs of observations, $(Y_{1}, \mbf{x}_{1}'),...,(Y_{n}, \mbf{x}_{n}')$ where $Y_{i}$ are observed count data and $\mbf{x}_{i}\in \mR^{p}$ are associated predictors. Let $\bX$ be an $n\times p$ matrix, the $i$th row vector of which is  $\mbf{x}_{i}'$. Define an $n\times p$ matrix $\bD:=((d_{ij}))$, $1\le i\le n$, $1\le j\le p$, where $d_{ij}$'s are the real-valued weights used to define the distance function in (\ref{eq:distance_function_regression}). It should be admitted that the following assumptions have a root in \cite{Koul2002} and \cite{Kim2026}.
\begin{itemize}
\item[(\tbf{a.\addtocounter{Qcounter}{1}\theQcounter})] Let $\textbf{B}$ denote an $n\times n$ symmetric, positive definite matrix. Then, $\bX'\textbf{B}\bX$ is nonsingular. In addition, there exists a $p\times p$ nonsingular matrix $\bA$ such that $\bA = (\bX'\textbf{B}\bX)^{-1/2}$.
\item[(\tbf{a.\addtocounter{Qcounter}{1}\theQcounter})] For all $1\le j\le p$, $\sum_{i=1}^{n}d_{ij}^{2}=1$, and $\max_{1\le i\le n}d_{ij}=o(1)$.
\item[(\tbf{a.\addtocounter{Qcounter}{1}\theQcounter})] Let $\mbf{c}_{ni}:=\bA \mbf{x}_{i}$ for $1\le i\le n$. Then $\max_{1\le i\le n}\|\mbf{c}_{ni}\|=o(1)$.
\item[(\tbf{a.\addtocounter{Qcounter}{1}\theQcounter})] For $1\le j\le n$, $\sti \|d_{ij}\mbf{c}_{ni}\| = O(1)$.
\item[(\tbf{a.\addtocounter{Qcounter}{1}\theQcounter})]
Let $a\vee b:=\max(a,b)$ for real values $a,b\in \mR$,  and let $\la_{i}^{0}=e^{\bxi'\B_{0}}$. Then $\max_{1\le i\le n}\{\la_{i}^{0}\vee \la_{i}:\,\B\in \cN_{b}(\B_{0})\}=O(1)$.
\item[(\tbf{a.\addtocounter{Qcounter}{1}\theQcounter})] Recall $\partial H(\cdot,\la;\cF)$ in (\ref{eq:gk_definition}), and let $g_{i}(\cdot,\B;\cF):=\la_{i}\partial H(\cdot,\la_{i};\cF)$. Let $\bG_{n}$ be an $n\times n$ diagonal matrix whose $i$th diagonal entry is $g_{i}$. Then a $p\times p$ matrix $\G_{n}:=\bD'\bG_{n}\bX\bA$ is nonsingular.
\item[(\tbf{a.\addtocounter{Qcounter}{1}\theQcounter})] Let $\mbf{d}_{i}'$ denote the $i$th row vector of $\bD$. Then, for all unit vectors $\mbf{e}\in \mR^{p}$,  either $\mbf{d}_{i}'\mbf{e}\mbf{x}_{i}'\bA\mbf{e}\ge 0$ or $\mbf{d}_{i}'\mbf{e}\mbf{x}_{i}'\bA\mbf{e}\le 0$ holds true for all $1\le i\le n$.
\item[(\tbf{a.\addtocounter{Qcounter}{1}\theQcounter})] For a unit vector $\mbf{e}\in \mathbb{R}^{p}$, let $k_{n}(\mbf{e}):=\mbf{e}'\G_{n}\mbf{e}$. Then there exists an $\alpha>0$ such that
    \benn
    \liminf_{n}\big\{\inf\{k_{n}(\mbf{e}):\mbf{e}\in \mathbb{R}^{p}\}\big\}\ge \alpha.
    \eenn
\end{itemize}
Based on the above assumptions, we will derive the MD estimator and its asymptotic properties in the next section.
\begin{rem}
The literature on MD estimation assumes the nonsingularity of $\bX'\bX$ and
\benn
\max_{1\le i\le n}\bxi'(\bX'\bX)^{-1}\bxi=o(1),
\eenn
which is called the ``Noether condition" of the design matrix. Note that \tbf{(a.1)} resembles the first assumption of the Noether condition, which is not a coincidence, while the second assumption is equivalent to \tbf{(a.3)}. In the setup of the continuous $Y_{i}$, the assumption that $Y_{i}$ are iid is common, and hence, if $\textbf{B}$ is any diagonal matrix whose entry is a derivative of distribution function (i.e., density function), then $\bX'\textbf{B}\bX$ will be reduced to a form of $\bX'\bX$ multiplied by the density function. In this study, we will encounter $\textbf{B}$ whose diagonal entries are derivatives of the pmf's with respect to different rate parameters; $\bG_{n}$ will be such an example. Thus, \tbf{(a.1)} and \tbf{(a.3)} imply the analogue of the Noether condition.
\end{rem}
\begin{rem}
In the literature on MD estimation for continuous probability distributions, additional assumptions about the density function are required. For example, for the probability density function $l(x)$, the following assumption is typical: $\int l^r(x) d\cM(x)<\iny$ for $r=1,2$, where $\cM(x)$ is an integrating measure. When the new approaches of the MD method -- using a pmf or df in the distance function -- is applied for the Poisson distribution, the integral and $l(x)$ are replaced by their discrete counterparts, namely, the sum and the derivative of the pmf or df with respect to the parameter, respectively. More importantly, the finiteness of the sum of derivatives, which plays a crucial role in the proof of the asymptotic normality, should be checked. Fortunately, due to those useful properties shown in (\ref{eq:gk_bdd}), we don't need such an assumption.
\end{rem}

\subsection{MD estimation for the Poisson regression}\label{sec:MD_regression_estimator}
To determine whether the ULAQ conditions for the distance function $\cL$ are satisfied, we first specify  $\mbf{S}_{n}$ and $\bW_{n}$ in \tbf{(U.1)}. Let $\cW_{j}(k,\B)$ denote the summand of $\cL$ in (\ref{eq:distance_function_regression}), that is,
\benn
\cW_{j}(k,\B;\cF):=\sti d_{ij}\{\cI(Y_{i},k; \cF)- H_{i}(k,\B;\cF)\}.
\eenn
Next define the following:
\benr
\mbf{S}_{n}(\B;\cF) &:=& -\stj\stk \sti \cW_{j}(k,\B;\cF)d_{ij}  \bqi(k,\B;\cF),\label{eq:Sn}\\
\bW_{n}(\B;\cF) &:=& \stj\stk\sum_{h=1}^{n}\sti d_{ij}d_{hj}\mbf{q}_{i}(k,\B;\cF)\mbf{q}_{h}'(k,\B;\cF),\nonumber\\
\cQ(\B;\cF) &:=& \cL(\B_{0};\cF) + 2(\B-\B_{0})'\mbf{S}_{n}(\B_{0};\cF)+(\B-\B_{0})'\bW_{n}(\B_{0};\cF)(\B-\B_{0}),\nonumber
\eenr
where $\mbf{q}_{i}(k,\B;\cF):=\partial H_{i}(k,\B;\cF)/\partial \B$. Note that $\mbf{q}_{i}(k,\B)=g_{i}(k,\B)\bxi$ where $g_{i}$ are from the assumption \tbf{(a.6)}. Recall $\cN_{b}(\B_{0})=\{\B\in \mathbb{R}^{p}:\,\|\bA^{-1}(\B-\B_{0})\|\le b\}$.
To prove the first ULAQ condition, we need the following lemma.
\begin{lem}\lel{lem:mu}
For $0< b<\iny$,
\benn
\sup_{\B\in \cN_{b}(\B_{0})}\stj\stk\left[ \sti d_{ij}\left\{ H_{i}(k,\B;\cF)-H_{i}(k,\B;\cF) -  (\B-\B_{0})'\mbf{q}_{i}(k,\B_{0};\cF) \right\} \right]^2 = o(1).
\eenn
\end{lem}
\begin{proof}
We will prove the claim for $\cF=f$ only since the proof of the other case will be almost identical. Rewrite the claim as
\ben\label{eq:lem_mu}
\sup_{\B\in \cN_{b}(\B_{0})}\stj\stk\left[ \sti d_{ij}\left\{ f_{i}(k;\B)-f_{i}(k;\B_{0}) -  (\B-\B_{0})'\mbf{q}_{i}(k,\B_{0};f) \right\} \right]^2 = o(1).
\een
Let $\mbf{u}:=\bA^{-1}(\B-\B_{0})\in \mR^{p}$. Recall $\mbf{c}_{ni}=\bA \mbf{x}_{i}\in \mR^{p}$, $1\le i\le n$. Observe that the mean value theorem (MVT) after replacing $\bqi$ with $g_{i}\bxi$ will yield
\benn
f_{i}(k;\B)-f_{i}(k;\B_{0}) -  (\B-\B_{0})'\mbf{q}_{i}(k,\B_{0};f)
= \mbf{u}'\mbf{c}_{ni}[g_{i}(k,\wt{\B};f)-g_{i}(k,\B_{0};f)],
\eenn
where $\wt{\B}=c\B_{0}+(1-c)\B$ for some $c\in (0,1)$. Let $\la_{i}^{0}=\bxi'\B_{0}$ and  $\wt{\la}_{i}=\bxi'\wt{\B}$. Note that $\partial g_{i}(k,\B;f)/\partial \B=(\la_{i}\partial H(k,\la_{i};f)+\la_{i}^2\partial^2 H(k,\la_{i};f))\bxi$, and hence, another application of MVT implies that with $\la_{i}^{*}$ being between $\la_{i}^{0}$ and $\wt{\la}_{i}$,
\benrr
\stk|g_{i}(k,\wt{\B};f)-g_{i}(k,\B_{0};f)|^2
&\le&\stk |\mbf{x}_{i}'(\wh{\B}-\B_{0})|^2\cdot |\la_{i*}^{2}\partial^2 H(k,\la_{i*};f)+\la_{i*}\partial H(k,\la_{i*};f)|^2,\\
&\le&\|\mbf{u}\|^2\|\mbf{c}_{ni}\|^2\left(2\la_{i*}^{4}\stk|\partial^2 H(k,\la_{i*};f)|^2+2\la_{i*}^{2}\stk |\partial H(k,\la_{i*};f)|^2\right),
\eenrr
where the first inequality follows from the Cauchy-Schwarz (CS) inequality, while the fact that $(a+b)^2\le 2(a^2+b^2)$ for $a,b\in \mR$ and $\|\wh{\B}-\B_{0}\|^2\le \|\B-\B_{0}\|^2$ readily implies the second inequality. Therefore,
\benrr
\textrm{the supremand in (\ref{eq:lem_mu})}
&\le&  \stj \stk\left[ \sti |d_{ij}\mbf{u}'\mbf{c}_{ni}|\cdot |g_{i}(k,\wt{\B};f)-g_{i}(k,\B_{0};f)| \right]^{2},\\
&\le& 2 p \|\bu\|^4 \left(\max_{1\le i\le n}\|\mbf{c}_{i}\|\right)^2
\left(\sti \|d_{ij}\mbf{c}_{ni}\|\right)^2\\
&&\qquad \times
\max_{1\le i\le n}\sup_{\la\in [\la_{i}^{0}, \la_{i}] }\left(\la^{4}\stk|\partial^2 H(k,\la;f)|^2+\la^{2}\stk |\partial H(k,\la;f)|^2\right),
\eenrr
where (\ref{eq:gk_bdd}) will imply that the last term of the second line is bounded, and hence, (\tbf{a.3}) and (\tbf{a.4}) with $\|\bu\|\le b$ will complete the proof of the lemma.
\end{proof}
\noi
When encountering any proofs involving $\cF$ in sequel, we will prove the case of $\cF=f$ only for the same reason as mentioned earlier. After stating a finding with full notations including $\cF$, we omit $\cF$ from all variables during the proof of the claim, unless specified otherwise; we write, e.g., $\cL(\B;\cF)$ and $g_{i}(k,\B;\cF)$ as $\cL(\B)$ and $g_{i}(k, \B)$, respectively, in the proof. The next theorem demonstrates that the first ULAQ condition is indeed satisfied.
\begin{thm}\label{thm:L_Q}
Assume (\tbf{a.1})-(\tbf{a.6}). Then, the distance function $\cL$ in (\ref{eq:distance_function_regression}) satisfies \textbf{(U.1)}, that is, for any $0<b<\iny$,
\benn
   \mathbb{E}\Big(\sup_{\B\in \cN_{b}(\B_{0})}|\cL(\B;\cF)- \cQ(\B;\cF)|\Big)=o(1).
\eenn
\end{thm}

\begin{proof}
Let $\mbf{u}=\bA^{-1}(\B-\B_{0})$ with $\|\mbf{u}\|\le b<\iny$. Note that $\cL$ and $\cQ$ can be rewritten in the following quadratic forms
\small{
\benrr
\cL(\B) &=& \stj\stk \Bigg[\Big\{\cW_{j}(k,\B_{0}) - (\B-\B_{0})'\sti d_{ij}\bqi(k,\B_{0})   \Big\} - \left. \sti d_{ij}\Big\{ f_{i}(k;\B)-f_{i}(k;\B_{0}) - (\B-\B_{0})'\bqi(k,\B_{0}) \Big\} \right]^{2},
\eenrr
}
\normalsize{and}
\benrr
\cQ(\B) &=& \stj\stk\left[\Big\{\cW_{j}(k,\B_{0}) - (\B-\B_{0})'\sti d_{ij}\bqi(k,\B_{0})   \Big\}\right]^2. %
\eenrr
\ifnum \value{show}=1{
Note that
\benn
\mathbb{E} \|\W(\B_{0})\|^{2} = \sum_{j=1}^{J}\stk d_{kj}^2p_{k}(\B)(1-p_{k}(\B))\le \sum_{j=1}^{J} \stk \le J. d_{kj}^2
\eenn
}\fi
Note that
\benn
\stk\mE \cW_{j}^{2}=\stk \sti d_{ij}^{2}f_{i}(k)[1-f_{i}(k)]\le \sti d_{ij}^{2}\stk f_{i}(k)=1,
\eenn
where the first equality follows from the independence assumption, $0\le f_{i}(k)\le 1$ implies the inequality, and the assumption \tbf{(a.2)} implies the last equality. Consequently,
\ben\label{eq:E_Wj}
\stj\stk \{\cW_{j}(k, \B_{0})\}^2=O_{p}(1).
\een
Next, observe that
\benr
\stj\stk\Big\{(\B-\B_{0})'\sti d_{ij}\bqi(k,\B_{0})\Big\}^2
&=&  \stj \stk \left|\mbf{u}'\sti d_{ij} \bA\bxi g_{i}(k,\B_{0})\right|^2,  \label{eq:O1}\\
&\le& p b^2 \stj \left(\sti \| d_{ij}\mbf{c}_{ni}\|\right)^2 \left(\max_{1\le i\le n}\stk \la_{i}^{2} |\partial H(k,\la_{i};f)|^{2} \right) =O(1), \nonumber
\eenr
where the last equality immediately follows from (\tbf{a.4}) and (\ref{eq:gk_bdd}). In view of Lemma \ref{lem:mu}, (\ref{eq:E_Wj}) and (\ref{eq:O1}), expanding the quadratic expression of $\cL$, subtracting $\cQ$ from it, and applying the CS inequality to the cross product term will complete the proof of the theorem.
\end{proof}
While Theorem \ref{thm:L_Q} ascertains (\tbf{U.1}), the other two ULAQ conditions are still unverified. To show that $\cL$ indeed satisfies (\tbf{U.2}) and (\tbf{U.3}), the following lemma is required.
\begin{lem}\label{lem:UALQ23}
In addition to the assumptions in Theorem \ref{thm:L_Q},
suppose that \tbf{(a.7)} and \tbf{(a.8)} hold. Then, the other ULAQ conditions are also satisfied.
\end{lem}
\begin{proof}
The proof of the lemma will be very similar to that of Lemma 3 from \cite{Kim2026}, and hence, we will only sketch the proof very briefly. To begin with, (\tbf{a.3}) and the independence of $Y_{i}$'s imply $\mE|\cL(\B_{0})|<\iny$, which, in turn, implies (\tbf{U.2}) by the Chebyshev's inequality. For some $\bu\in \mR^{p}$, define
\benn
V_{j}(\bu):=\stk \cW_{j}(k,\B_{0}+\bA\bu)l(k), \quad \wh{V}_{j}(\bu):=\stk \{\cW_{j}(k,\B_{0})+\bu\G_{n}'(k,\B_{0})\}l(k),
\eenn
where $l:\{0\}\cup\mathbb{N}\ra \mR$ such that $\stk l^{2}(k)<\iny$; for example, a pmf of a Poisson distribution with any rate parameter -- or any other discrete distribution -- can be used for $l$. Subsequently, define $\mbf{V}(\bu):=(V_{1},...,V_{p})'\in \mR^{p}$ and $\wh{\mbf{V}}(\bu):=(\wh{V}_{1},...,\wh{V}_{p})'\in \mR^{p}$. Then,
\ben\label{eq:V_Vhat}
\sup_{\|\bu\|\le b}\|\wh{\mbf{V}}(\bu)- \mbf{V}(\bu)\|=o(1)
\een
will follow from Lemma \ref{lem:mu} and $\stk l^{2}(k)<\iny$ after application of the CS inequality. In view of the assumptions (\tbf{a.7}) and  (\tbf{a.8}), it can be shown that the monotonicity of both $\wh{\mbf{V}}(\bu)$ and $\mbf{V}(\bu)$ in $\|\bu\|$. Finally, as done in \cite{Koul2002} and \cite{Kim2026}, the monotonicity of $\mbf{V}$ and $\wh{\mbf{V}}$ combined with (\ref{eq:V_Vhat}) will yield (\tbf{U3}), thereby completing the proof of the lemma.
\end{proof}

The next lemma shows the asymptotic normality of $\bA \bSn$, which is required for that of the MD estimator. Recall $\G_{n}(\B)=\bD'\bG_{n}\bX\bA$ from the assumption \tbf{(a.5)}. Define $\wt{\G}_{n}(\B):=\stk \G_{n}'(k;\B)\G_{n}(k;\B)$. Let $\W(k,\B):=(\cW_{1},...,\cW_{p})'\in\mR^{p}$. Note that $\bA\bSn$ and $\bA\bW_{n}\bA$ in the following lemma can be expressed using these matrix and vector: $\bA\bSn = \stk \G_{n}'\W$ and $\bA\bW_{n}\bA=\wt{\G}_{n}$. Let  $\bO_{n}(\B):=\stk \bGa_{n}'\bD' \bP_{n}\bD\bGa_{n}$, where $\bP_{n}(k;\B)$ is an  $n\times n$ diagonal matrix whose $i$th entry is $f_{i}(k;\B)\{1-f_{i}(k;\B)\}$.
\begin{lem}\label{lem:Asym_Convergence_Sn_Regression}
Assume that $\wt{\G}_{n}(\B_{0})$ is positive definite, and \benn
\lim_{n\ra\iny} \wt{\G}_{n}(\B_{0}) = \wt{\G}(\B_{0}).
\eenn
Then, $\cB_{n}:=\bA\bW_{n}\bA$ converges to $\wt{\G}(\B_{0})$, and
\benn
\bO_{n}^{-1/2}\bA\mbf{S}_{n}(\B_{0})\Rightarrow_{\cD}N(\mathbf{0}_{p\times 1}, \mathbf{I}_{p\times p}),
\eenn
as $n$ increases to $\iny$.
\end{lem}
\begin{proof}
The convergence of $\bA\bW_{n}\bA$ is trivial. Let
$\mbf{\gamma}_{j}(k,\B)$ denote the $j$th column vector of $\bGa_{n}'(k,\B)$. Note that
$\mbf{\gamma}_{j}(k,\B_{0})=\sti d_{ij}\mbf{c}_{ni}g_{i}(k;\B_{0})$. Hence, \tbf{(a.4)}, \tbf{(a.5)}, and (\ref{eq:gk_bdd}) will imply
\ben\label{eq:gamma_j}
\stk\|\mbf{\gamma}_{j}(k,\B_{0})\|<\iny, \quad \forall\,j=1,2,...,p.
\een

Next, for $\mbf{a}=(a_{1},...,a_{p})'\in \mathbb{R}^{p}$, rewrite $\mbf{a}'\bA\mbf{S}_{n}$
\benrr
\mbf{a}'\bA\mbf{S}_{n}(\B_{0})&=& \sti \stj d_{ij}  \sum_{l=1}^{p}a_{l}\stk \mbf{\gamma}_{j}(k,\B_{0})\Big\{\textrm{I}(Y_{i}=k)-f_{i}(k;\B_{0}) \Big\}\\
&=& \sti \eta_{i}(\B_{0}),\qquad (say).
\eenrr
Observe that for the bounded $\mbf{a}$
\ben\label{eq:eta_bdd}
|\eta_{i}(\B_{0})|\le \|\mbf{a}\|\stj\stk\|\mbf{\gamma}_{j}(k,\B_{0})\|<\iny.
\een
It is clear to see that $\mathbb{E}[\eta_{i}(\B_{0})]=0$ for all $1\le i\le n$. To conserve space, we occasionally drop $\B_{0}$ from variables if they contain it as an argument. For example, we write $\eta_{i}(\B_{0})$ and $\mbf{\gamma}_{j}(\cdot, \B_{0})$ as $\eta_{i}$ and $\mbf{\gamma}_{j}(\cdot)$, respectively.  Let $\si_{i}^{2}:=\mathbb{E}(\eta_{i}^2)$ and $\tau_{n}^{2}:=\sti \si_{i}^{2}$. Definitely, both $\si_{i}^{2}$ and $\tau_{n}^{2}$ are functions of $\B_{0}$. Hence, for any $\epsilon>0$.
\benrr
\tau_{n}^{-2}\sti \mathbb{E}(\eta_{i}^{2}:|\eta_{i}|\ge\epsilon\tau_{n})&\le& C\tau_{n}^{-2}\max_{1\le j\le p}\max_{1\le i\le n} d_{ij}^2 \left(\stj\stk\|\mbf{\gamma}_{j}(k)\|\right)^2 \sti \mathbb{P}(|\eta_{i}|\ge\epsilon\tau_{n})\\
&\le& C\epsilon^{-2}\tau_{n}^{-2} \max_{1\le j\le p}\max_{1\le i\le n} d_{ij}^2  \left(\stj\stk\|\mbf{\gamma}_{j}(k)\|\right)^2  \longrightarrow 0,
\eenrr
where the first inequality follows from (\ref{eq:eta_bdd}), and the second inequality follows after application of the Chevyshev inequality to the summand of the last term in the first line, whereas the convergence to 0 follows from \tbf{(a.2)} and (\ref{eq:gamma_j}), thereby showing that the Lindeberg-Feller (LF) condition is satisfied. Recall $\bO_{n}(\B)$ and note that
\benn
\tau_{n}^{2}(\B_{0})=\mbf{a}'\mE\{\bA\bSn(\B_{0})\bSn'(\B_{0})\bA\}\mbf{a}=\mbf{a}'\bO_{n}(\B_{0})\mbf{a}.
\eenn
Thus, the claim follows from the Cramer-Wold device with the LF condition, thereby completing the proof of the theorem.
\end{proof}
\begin{rem}
In the references of MD estimation in the continuous setup, the analogue of
(\ref{eq:gamma_j}) -- $g_{i}$ being replaced with the density function of the continuous variable -- is assumed in order to bound $\eta_{i}$ in (\ref{eq:eta_bdd}); see, e.g., the assumption (h) of \citet[p,\,174]{Koul2002}. Unlike the references, (\ref{eq:eta_bdd}) can be established without (\ref{eq:gamma_j}) in this study. However, the assumption of non-singularity of $\G_{n}$ is still required during this study.
\end{rem}
We conclude this section by stating the main result of this study: the asymptotic normality of the MD estimator. 
\begin{thm}\label{thm:asymp_Normal_Poisson_Regression}
Suppose the assumptions in Theorem \ref{thm:asymp_Normal_Poisson_Regression} and Lemma \ref{lem:Asym_Convergence_Sn_Regression} hold. Let $\wt{\G}_{n}$ and $\bO_{n}$ be as in Lemma \ref{lem:Thm541}. Then
the MD estimator $\wh{\B}$ asymptotically follows the normal distribution, that is,
\benn
 \S_{n}^{-1/2} \bA^{-1}(\wh{\B}-\B_{0}) \Rightarrow_{\cD} N(\mathbf{0}_{p\times 1}, \mathbf{I}_{p\times p}),
\eenn
where $\S_{n}:=\wt{\G}_{n}^{-1}(\B_{0})\bO_{n}(\B_{0})\wt{\G}_{n}^{-1}(\B_{0})$.
\end{thm}
\begin{proof}
Theorem \ref{thm:L_Q} and Lemma \ref{lem:UALQ23} ensure that the ULAQ conditions are met, and hence, the asymptotic normality of the MD estimator immediately follows from Lemmas  \ref{lem:Thm541} and \ref{lem:Asym_Convergence_Sn_Regression}.
\end{proof}
\begin{rem}
Direct implication of the previous Lemma \ref{lem:Asym_Convergence_Sn_Regression} and Theorem \ref{thm:asymp_Normal_Poisson_Regression} is the provision of statistical inference about the unknown $\B_{0}$ -- such as a hypothesis test -- in addition to the asymptotic distribution; the results of the lemma provides a Score-type test, whereas those of the theorem can be a starting point to develop a Wald-type test.
\end{rem}

\subsection{Robustness of MD estimator}\label{sec:MD_robustness}
Note that minimizing the distance function in (\ref{eq:distance_function_regression}) is equivalent solving
\benn
\sti \phi(Y_{i}, \bxi;\B) = 0,
\eenn
where
\benn
\phi(Y_{i}, \bxi;\B) = \stk \bqi(k,\B)\stj d_{ij} \cW_{j}(k,\B).
\eenn
Taking a partial derivative of $\phi$ with respect to $\B$ yields
\benn
\frac{\partial \phi(Y_{i}, \bxi;\B)}{\partial \B} = \stk \left[\bqi(k,\B)\frac{\cW_{i}^{*}(k, \B)}{\partial \B}+\frac{\bqi(k,\B)}{\partial \B}\cW_{i}^{*}(k, \B)\right],
\eenn
where $\cW_{i}^{*}(k,\B):= \stj d_{ij}\cW_{j}(k,\B)$.
Recall that $\mE(\cW_{j})=0$ for all $1\le j\le p$. Hence,  $\mE(\cW_{i}^{*})=0$ is straightforward, thereby implying
\benn
\mE\left[\frac{\partial \phi(Y_{i}, \bxi;\B)}{\partial \B}\right] = -\stk \sum_{h=1}^{n}d_{hi}^{*} g_{i}(k,\B)g_{h}(k,\B)\bxi\mbf{x}_{h}'.
\eenn
Let $\textrm{IF}(Y_{i},\bxi;\B)$ denote an influence function of the MD estimator when $(Y_{i}, \bxi)$ are the observed data points. This influence function measures the robustness of the MD estimator, especially in the presence of the huge outlier $\bxi$. Then, the direct application of the formula (2.3.5) from \citet[p.\,\,101]{Hampel1986} yields
\benrr
\textrm{IF}(Y_{i},\bxi;\B) &=& -\left(\mE\left[\frac{\partial \phi(Y_{i}, \bxi;\B)}{\partial \B}\right] \right)^{-1} \phi(Y_{i}, \bxi;\B)\\
&=& \left(\stk \sum_{h=1}^{n}d_{hi}^{*} g_{i}(k,\B)g_{h}(k,\B)\bxi\mbf{x}_{h}'\right)^{-1}
\left(\sum_{k=0}^{\iny}g_{i}(k,\B)\bxi \cW_{i}^{*}(k,\B)\right).
\eenrr
Consider a scalar $x_{i}\in \mR$ and $\la_{i}=x_{i}\beta$. Then,
\benrr
\textrm{IF}(Y_{i},x_{i};\beta) &=& \frac{\sum_{k=0}^{\iny}g_{i}(k,\b) \cW_{i}^{*}(k,\b)}{\stk \sum_{h=1}^{n}d_{hi}^{*} g_{i}(k,\b)g_{h}(k,\b)x_{h}},\\
&=& \frac{\sum_{k=0}^{\iny} \partial f(k;\la_{i}) \cW_{i}^{*}(k,\B)}{ \sum_{h\neq i}d_{hi}^{*} \stk \partial f(k;\la_{i}) g_{h}(k,\b)x_{h}+d_{ii}^{*}x_{i}\stk \partial f(k;\la_{i}) g_{i}(k, \beta) },\\
\eenrr
where $|\sum_{k=0}^{\iny}\partial f(k;\la_{i}) \cW_{i}^{*}(k,\b)|<\iny$ immediately follows from (\ref{eq:gk_bdd}). Note that
\benrr
x_{i}\stk \partial f(k;\la_{i}) g_{i}(k,\b) &=& \frac{x_{i}}{\la_{i}}\stk \left( \frac{\la_{i}^{k}e^{-\la_{i}}}{k!}\right)^{2}(k-\la_{i})^2\\
&\le&\frac{2x_{i}}{\la_{i}}\left[\stk \left( \frac{\la_{i}^{k}e^{-\la_{i}}}{(k-1)!}\right)^{2}+\stk \left( \frac{\la_{i}^{k+1}e^{-\la_{i}}}{k!}\right)^{2}\right],
\eenrr
where the last inequality follows from $(a-b)^2\le 2(a^2+b^2)$. Note that the ratio test implies that the two series converge, while $x_{i}/\la_{i}=x_{i}/e^{x_{i}\beta}$ approaches 0 as $x_{i}$ increases. Thus, the influence function $\textrm{IF}(Y_{i},x_{i};\b)$ is bounded unless the first term -- which is unaffected by $x_{i}$ -- of the denominator stays away from 0, thereby implying the impact of the outlier is limited.

\subsection{Bias reduction for the Poisson regression}\label{sec:MD_bias_reduction}
In a series of recent research studies, it has been observed that the MD estimator often exhibits relatively large bias, thereby countervailing its efficiency achieved by the circumspect choice of $\bD$ and integrating measures; see simulation studies in \cite{Kim2018}, \cite{Kim2020}, and \cite{Kim2026}. Thus, reducing the bias of the MD estimator to some extent will further consolidate its superiority over others. To this end, we will investigate whether bias reduction of the MD estimator is conceivable.

In the statistical literature on maximum likelihood (ML) estimation, bias reduction has been a popular topic, and a series of studies, ranging from the foundational work by \cite{Cox1968} to the most recent work by \cite{Kosmidis2021}, have discussed bias reduction of the ML estimator. The most common and popular technique was expanding of the score function to the first or second order and solving the resulting equation in terms of bias. Emulating this approach, we apply the Taylor expansion to the counterpart of the ML's score function, namely $\partial\cL(\B)/\partial \B$, and obtain
\ben\label{eq:taylor_l}
\left. \frac{\partial\cL(\B)}{\partial \B}  \right|_{\B=\wh{\B}}
\approx \left. \frac{\partial\cL(\B)}{\partial \B}  \right|_{\B=\B_{0}}+(\wh{\B}-\B_{0})'\left. \frac{\partial^{2}\cL(\B)}{\partial \B\partial \B'}  \right|_{\B=\B_{0}}.
\een
Let $\bR(\cdot,\B):=\bA^{-1}\G_{n}'(\cdot,\B)$ and $\mbf{R}_{j}$ denote its $j$th column. Note that $\mbf{R}_{j}(k,\B)=\stj d_{ij}g_{i}(k,\B)\bxi$ and $\bSn=\stk \stj\mbf{R}_{j}(k,\B)\cW_{j}(k,\B)$. Therefore, we have
\benn
\mE \left[\left. \frac{\partial\cL(\B)}{\partial \B}  \right|_{\B=\B_{0}}\right] = -2\mE\{\bSn(\B_{0})\} = -2\stj \mbf{R}_{j}(k,\B) \mE\{\cW_{j}(k,\B_{0})\} = \mbf{0}_{p\times 1},
\eenn
where the last equality follows from $\mE\{\cW_{j}(k,\B_{0})\}=0$. Since the left-hand side of the equation in (\ref{eq:taylor_l}) will be $\mbf{0}_{p\times 1}$, with the above equation, taking an expectation on both-hand sides of the equation yields
\ben\label{eq:bias_reduction}
\mE \left[(\wh{\B}-\B_{0})'\left. \frac{\partial^{2}\cL(\B)}{\partial \partial\B\partial\B'}  \right|_{\B=\B_{0}}  \right] \approx 0.
\een
Thus, we will obtain the approximate bias of the MD estimator by solving the above equation. Let $\dot{g}_{i}(k,\B):=\partial g_{i}(k,\B)/\partial \la_{i}$ to conserve space. Observe that $\partial \mbf{R}_{j}/\partial \B= \sti d_{ij}\la_{i} \dot{g}_{i}\bxi\bxi'$, and hence,
\benn
\cW_{j}\frac{\partial \mbf{R}_{j}}{\partial \B} = \sti d_{ij} \la_{i} \dot{g}_{i}\cW_{j}\bxi\bxi'= \bX'\bL_{j}\bX,
\eenn
where $\bL_{j}$ is an $n\times n$ diagonal matrix whose $i$th entry is $d_{ij}\la_{i}\dot{g}_{i}\cW_{j}$.
Note that
\benrr
\frac{\partial^{2} \cL(\B)}{\partial \B\partial \B'}&=&-2\stk \frac{\partial }{\partial \B} \stj\mbf{R}_{j}(k,\B)\cW_{j}(k,\B), \\
&=& -2\stk\left[ \bX'\left(\stj\bL_{j}(k,\B)\right)\bX- \stj \mbf{R}_{j}(k,\B)\mbf{R}_{j}'(k,\B) \right],\\
&=&-2\bX'\bL^{*}(\B)\bX+2 \bW_{n}(\B),
\eenrr
where $\bL^{*}(\B):=\stk\stj\bL_{j}(k,\B)$; the second equality follows from $\partial \cW_{j}/\partial \B= \mbf{R}_{j}$, whereas the last equality is true since, with $\bqi$ being replaced with $g_{i}\bxi$, $\bW_{n}$ in (\ref{eq:Sn}) can be written as
\benn
\bW_{n} = \stk\stj\left(\sti d_{ij}g_{i}\bxi \right) \left(\sum_{h=1}^{n}d_{hj}g_{h}\mbf{x}_{h}'\right)
= \stk \stj \mbf{R}_{j}(k,\B)\mbf{R}_{j}'(k,\B).
\eenn
Therefore, plugging the second order derivative of $\cL$ into (\ref{eq:bias_reduction}) will yield
\benn
\mE\left[ (\wh{\B}-\B)' \frac{\partial^{2} \cL(\B)}{\partial \B\partial \B'}\right]=2\bW_{n}\mE(\wh{\B}-\B)-2\mE\left[(\wh{\B}-\B)'\bX'\bL^{*}(\B)\bX\right]\approx 0,
\eenn
where the equality is true since there is no randomness in $\bW_{n}$. Consequently, the approximate bias of the MD estimator has an analytic expression
\benrr
\mE(\wh{\B}-\B)&\approx& \bW_{n}^{-1}\mE\left[(\wh{\B}-\B)'\bX'\bL^{*}(\B)\bX\right],\\
&=& \bW_{n}^{-1}\stk\stj \mE\left[(\wh{\B}-\B)'\bX'\bL_{j}(k,\B)\bX\right],\\
&=& \bW_{n}^{-1}\stk\stj  \sti \left( d_{ij}\la_{i}\dot{g}_{i}(k,\B)\bxi\bxi'\mE\left[\cW_{j}(k,\B)(\wh{\B}-\B)\right]\right),
\eenrr
which implies that only task left is to find the $\mE\left[\cW_{j}(\cdot,\B)(\wh{\B}-\B)\right]$.

Recall  $\eta_{i}(k,\B):=\textrm{I}(Y_{i}=k)-f_{i}(k;\B),\,1\le i\le n$, and let $\mbf{\eta}(k, \B):=(\eta_{1},...,\eta_{n})'\in \mR^{n}$. Define an $n\times n$ matrix $\Upsilon(l,k,\B):=\mE\left[\mbf{\eta}(l,\B)\mbf{\eta}'(k,\B)\right]$, $1\le l,k\le n$. Note that $\Upsilon$ is an $n\times n$ diagonal matrix due to the independence of $Y_{i}$'s, where its $(i,i)$th diagonal entry will be; $f_{i}(k;\B)\{1-f_{i}(k;\B)\}$ if $l=k$; $-f_{i}(l;\B)f_{i}(k;\B)$, otherwise. Note that the asymptotic quadraticity of the distance function, in other words, the asymptotic approximation of $\cL$ by the quadratic function $\cQ$, will imply that
\benn
(\wh{\B}-\B)\approx -\bW_{n}^{-1}\bSn(\B) = -\bW_{n}^{-1}\sum_{l=0}\bR(l, \B)\W(l,\B).
\eenn
Let $\mbf{d}_{j}$ denote the $j$th column vector of $\bD$.
Thus, using $\W = \bD'\mbf{\eta}$ and $\cW_{j} = \mbf{\eta}'\mbf{d}_{j}$, we have
\benrr
\mE\left[\cW_{j}(k,\B)(\wh{\B}-\B)\right]&\approx&\bW_{n}^{-1}\sum_{l=0}^{\iny}\bR(l,\B)\mE\left[\cW_{j}(k,\B)\W(l,\B)\right],\\
&=& \bW_{n}^{-1}\left(\sum_{l=0}^{\iny}\bR(l,\B)\bD'\Upsilon(l,k,\B)\right)\mbf{d}_{j},\\
&=& \mbf{\upsilon}_{j}(k,\B),\quad (say).
\eenrr
Next, plugging $\mbf{\upsilon}_{j}$ into the above equation
will lead to
\benrr
\sti  d_{ij}\la_{i}\dot{g}_{i}(k,\B)\bxi\bxi'\mE\left[\cW_{j}(k,\B)(\wh{\B}-\B)\right]
&=& \left(\sti d_{ij}\la_{i}\dot{g}_{i}(k,\B)\bxi\bxi'\right)\mbf{\upsilon}_{j},\\
&=&\bX'\bL_{j}^{0}(k,\B)\bX\mbf{\upsilon}_{j},
\eenrr
where $\bL_{j}^{0}$ is an $n\times n$ diagonal matrix whose $i$th entry is $d_{ij}\la_{i}\dot{g}_{i}(k,\B)$. Note the difference between $\bL_{j}^{0}(k,\B)$ and $\bL_{j}(k,\B)$. Putting all together, we obtain the bias of the MD estimator
\benn
\mE(\wh{\B}-\B_{0}) \approx\Big[\bX'\bL^{*}(\B_{0})\bX\Big]^{-1} \left[
\stk\stj \bX'\bL_{j}^{0}(k,\B_{0})\bX\mbf{\upsilon}_{j}(k,\B_{0})\right].
\eenn



\section{Generalized Poisson regression}\label{Sec:GP_Regression}
\subsection{Generalized Poisson distribution}\label{sec:GP_distribution}
\cite{Consul} proposed the generalized Poisson (GP) distribution with two parameter $\la$ and $\phi$ whose pmf is
\benn
\mP(Y=k;\la,\vp) = \frac{\la}{k!}(\la+\vp k)^{k-1}e^{-(\la+\vp k)},
\eenn
where $Y$ is the count data taking integer values. Depending on whether $\phi$ takes a negative or positive value, the mean of the GP distribution can be smaller or larger than variance; when $\phi=0$, the GP distribution will be the regular Poisson distribution, and hence, its mean will the same as its variance again. To distinguish it from the different version of the GP distribution in the sequel, we call it the \textit{original GP} distribution.

Reparameterizing the Consul's GP pmf, \cite{Famoye} proposed another pmf, which is more suitable for a regression setup, that is,
\benn
\mP(Y=k;\la, \vp)= \left(\frac{\la}{1+\vp\la}\right)^{k}\frac{(1+\vp k)^{k-1}}{k!}e^{-\la(1+\vp k)/(1+\vp \la)}.
\eenn
Through the rest of this article, we refer to the reparameterized distribution and pmf as simply GP distribution and pmf, respectively. For the GP distribution, the mean and variance are $\la$ and $\la(1+\vp \la)^2$, respectively, and hence, the variance will become smaller than, the same as, or larger than the mean upon the sign of $\phi$. Note that its mean is parameterized only by $\la$, shedding a clue to why GP2 is more convenient for the GP regression. In this article, we will employ the GP distribution -- not the original GP one -- when applying the MD method to the GP regression in the next section.

Before focusing the GP regression intently, we will further investigate the pmf of the GP distribution and derive some features that are analogues of (\ref{eq:gk_bdd}) and turn out to be useful for some proofs in the next section. Let $f(k;\la, \vp)=\mP(Y=k;\la, \vp)$.
Let $\partial_{\la}f(k;\la, \vp)$ and $\partial_{\phi}f(k;\la, \vp)$ denote the first order derivative of $f$ with respect to $\la$ and $\vp$, being written as
\ben\label{eq:GP_pmf_derivative}
\partial_{\la}f(k;\la, \vp)= f(k;\la, \vp) \frac{k-\la}{\la(1+\vp\la)^2},\,\,\,
\partial_{\vp}f(k;\la, \vp)= f(k;\la, \vp) \frac{\la k - \la^{2}+k\la+ k\vp \la^2}{(1+\vp \la)^2}.
\een
Note that $\vp=0$ will reduce $\partial_{\la}f(k;\la, \vp)$ to that of (\ref{eq:gk_definition}). We have analogues of (\ref{eq:gk_bdd}), that is, for $r=1,2$,
\ben\label{eq:gk_bdd2}
\stk |\partial_{\la}f(k;\la, \vp)|^r=O(1),\,\,\stk |\partial_{\vp}f(k;\la, \vp)|^r=O(1).
\een
Furthermore, being bounded also holds for the second-order partial derivatives, that is,
\ben\label{eq:gk_bdd3}
\stk |\partial_{\la}^2 f(k;\la, \vp)|^r=O(1),\,\,\stk |\partial_{\vp}^2f(k;\la, \vp)|^r=O(1),\,\,\stk |\partial_{\la}\partial_{\vp}f(k;\la, \vp)|^r=O(1),
\een
where $\partial_{\la}\partial_{\vp}$ denote the
second-order mixed derivative. The proofs of (\ref{eq:gk_bdd2}) and (\ref{eq:gk_bdd3}) are very similar to that of (\ref{eq:gk_bdd}), albeit more complicated, and we do not include it here. For example, using (\ref{eq:GP_pmf_derivative}), it can be easily shown that
\benn
\stk |\partial_{\la}^2 f(k;\la, \vp)|\le \frac{1+3\vp\la}{\la(1+\phi\la)^3}.
\eenn
It can be shown that the analogues of (\ref{eq:gk_bdd2}) and (\ref{eq:gk_bdd3}) for $F$ hold true as in Section $\ref{Sec:MD_one_sample}$, even though the proof will be more complicated and much longer.

\subsection{Extension of MD estimation to GP regression}\label{Sec:MD_GP_regression}
When examining application of MD estimation to the GP regression in this section, we will observe that the change of dimensions of variables as a new dispersion parameter is added into the analysis. Thus, considering GP $f$ and $F$ concurrently for the distance function, and hence, including the notation $\cF$ in variables will leave a room for confusion. Thus, we will examine the distance function using $f$ and non-cumulative indicator function only and demonstrate that analogues of findings obtained on the setup of the original Poisson regression will continue to hold. The verification of the claim as to using $F$ and cumulative indicator function will be left to readers who are interested. 
  
Now assume that the count data $Y_{i},\,\,1\le i \le n$ are independently distributed with a GP $f(k;\la_{i},\vp_{i})$ where $\la_{i}$ and $\vp_{i}$ are affected by two predictors $\bxii\in \mR^{p_{1}}$ and $\bxiii\in \mR^{p_{2}}$, that is,
\benn
\la_{i}=e^{\bxii'\B},\quad \vp_{i}=\bxiii'\D.
\eenn
Note that $\vp_{i}$ is linearly associated with $\bxiii$, which allows $\vp_{i}$ to take negative values so that the dispersion can be smaller than the mean as explained in Section \ref{sec:GP_distribution}.
Let $\bTh_{i}=(\la_{i}, \vp_{i})'\in \mR^{2},\,1\le i\le n$. Also, let $\bth:=(\B,\D)'\in \mR^{p}$ where $p=p_{1}+p_{2}$. In sequel, rewrite $f(k;\la_{i},\vp_{i})=f_{i}(k;\bth)$ to denote the main parameter of interest is $\bth$. Now define an $p\times 2$ matrix $\bX_{i}$, which can be partitioned into a $2\times 2$ block
\benn
\bX_{i} = \left[
          \begin{array}{cc}
            \bxii &  \mbf{0}_{1} \\
            \mbf{0}_{2} & \bxiii \\
          \end{array}
        \right],
\eenn
where the dimensions of the two zero vectors match those of $\bxii$ and $\bxiii$, respectively. Next, define
\benn
g_{1i}(k,\bth):=\la_{i}\frac{\partial}{\partial \la_{i}} f_{i}(k;\bth),\,\,\,g_{2i}(k,\bth):=\frac{\partial}{\partial \vp_{i}} f_{i}(k;\bth),
\eenn
and $\bg_{i}(k,\bth):=(g_{1i}, g_{2i})'\in \mR^{2}$.
Observe that $\partial f_{i}(k;\bth)/\partial \bth= \bX_{i}\bg_{i}$. Finally, define a design matrix $\bX$ by stacking $\bX_{i}', 1\le i\le n$ and another matrix $\bG_{n}$ by diagonalizing $\bg_{i}, 1\le i\le n$. As a result, the dimensions of $\bX$ and $\bG_{n}$ are $2n\times p$ and $2n\times n$, respectively; note that $\bG_{n}$ is not square anymore, but it can be still partitioned into an $n\times n$ block.

At this juncture, several facts are worth mentioning. First, $\bX_{i}'$ and $\bg_{i}$ plays roles of $\bxi'$ and $g_{i}$ of the original Poisson regression setup. Second, the appearance of the new parameter $\D$ \textit{doubles} the dimension -- more precisely, the number of rows -- of $\bxi'$ and $g_{i}$ of the original Poisson regression setup;  for example, $\bxi'$ is an $1\times p$ row vector (or matrix), while $\bX_{i}'$ is an $2\times p$ matrix, thereby causing the dimension of the design matrix $\bX$ to increase also from $n\times p$ to $2n\times p$. The same fact holds for $\bg_{i}$ and $\bG_{n}$. Last but not least, there will be some variables whose dimension still doesn't change; $\bD$ is one of them with its dimension being $n\times p$. As a result, the $\G_{n}$ of the previous section can be expressed as the same, namely, $\G_{n}=\bD'\bG'\bX$. When we demonstrate the asymptotic properties of the MD estimator of the GP regression, we will recycle these variables.
Now we define a new distance function for the GP regression for $\bth=(\B', \D')'\in \mR^{p}$ as
\ben\label{eq:distance_function_regression_GP}
\cL(\bth)=\stj\stk\left[\sti d_{ij}\left\{ \textrm{I}(Y_{i}= k)-f_{i}(k;\bth) \right\}\right]^2,
\een
and the corresponding MD estimator $\wh{\bth}=(\wh{\B}', \wh{\D}')'$ will solve the following optimization problem
\benn
\cL(\wh{\bth}) = \inf_{\theta\in\mathbb{R}^{p}}\cL(\bth).
\eenn
Note that the asymptotic properties of the MD estimator of the GP regression still require the same ULAQ conditions but different assumptions than those of the previous sections, as another parameter $\D$ is added and causes the dimensional changes of some variables. Thus, we shall modify the assumptions accordingly, which implies that the assumption regarding variables keeping the same dimension will remain intact. For example, we will keep (\tbf{a.2}) since there is no change in $\bD$.

To begin with, define a new neighborhood $\cN_{b}(\bth_{0})=\{\bth=(\B,\D)'\in \mathbb{R}^{p}:\,\|\bA^{-1}(\bth-\bth_{0})\|\le b\}$, where $\bA$ is a $p\times p$ matrix that satisfies the assumption (\tbf{a.1})' below. Since we have a design matrix of a new dimension, we will replace (\tbf{a.1}) with the following assumption:
\begin{itemize}
\item[(\tbf{a.1})'] Let $\textbf{B}$ denote an $2n\times 2n$ symmetric, positive definite matrix. Then, $\bX'\textbf{B}\bX$ is nonsingular. In addition, there exists a $p\times p$ nonsingular matrix $\bA$ such that $\bA = (\bX'\textbf{B}\bX)^{-1/2}$.
\end{itemize}
Now we will partition $\bA$ into a $2\times 2$ block so that diagonal blocks are $p_{1}\times p_{1}$ and $p_{2}\times p_{2}$ matrices, denoted by $\bA_{1}$ and $\bA_{2}$, respectively. Regarding the changes in $\bxi$'s, we replace \tbf{(a.3)}, \tbf{(a.4)}, and \tbf{(a.5)} with
\begin{itemize}
\item[\tbf{(a.3)}'] Let $\mbf{c}_{1i}:=\bA_{1} \bxii$ and $\mbf{c}_{2i}:=\bA_{2} \bxiii$ for $1\le i\le n$. Then $\max_{1\le i\le n}\|\mbf{c}_{\ell i}\|=o(1)$ for $\ell=1,2$.
\item[\tbf{(a.4)}'] For $1\le j\le p$, $\stk \|d_{ij}\mbf{c}_{\ell i}\| = O(1)$ for $\ell=1,2$.
\item[\tbf{(a.5)}'] Let $\bTh_{i}^{0}:=(\la_{i}^{0}, \vp_{i}^{0})$ where $\la_{i}^{0}=e^{\bxii'\B_{0}}$ and $\vp_{i}^{0}=\bxiii'\D_{0}$. Then $\max_{1\le i\le n}\{\|\bTh_{i}^{0}\|\vee \|\bTh_{i}\|:\,\bth\in \cN_{b}(\bth_{0})\}=O(1)$.
\end{itemize}
\begin{rem}
\tbf{(a.5)}' immediately implies \tbf{(a.5)} and $\max_{1\le i\le n}\{|\vp_{i}^{0}|\vee |\vp_{i}|\}<\iny$.
\end{rem}
Note that $\bA\bX_{i}$ can be expressed as
\benn
\bA\bX_{i} = \left[
          \begin{array}{cc}
            \bA_{1}\bxii &  \mbf{0} \\
            \mbf{0} & \bA_{2}\bxiii \\
          \end{array}
        \right],
\eenn
and hence, for any $\bu:=(\bu_{1}',\bu_{2}')=\bA^{-1}(\bth-\bth_{0})\in \mR^{p}$ with $\bu_{1}\in \mR^{p_{1}}$ and $\bu_{2}\in \mR^{p_{2}}$, we have
\ben\label{eq:uAX}
(\bth-\bth_{0})'\bX_{i}\bg_{i}(k,\bth)= \bu_{1}'\mbf{c}_{1i}g_{1i}(k,\bth)+\bu_{2}'\mbf{c}_{2i}g_{2i}(k,\bth).
\een
Note that for any $\bth=(\B',\D')'\in \cN_{b}(\bth_{0})$ we have $\|\bu_{1}\|, \|\bu_{2}\|\le b$. With these new assumptions, we can prove the analogues of all lemmas and theorems from Section \ref{Sec:MD_regression}: Lemmas \ref{lem:mu} and \ref{lem:Asym_Convergence_Sn_Regression} and Theorems \ref{thm:L_Q} and \ref{thm:asymp_Normal_Poisson_Regression}. Since the analogues can be shown in a similar way, we will prove only the analogue of Lemma \ref{lem:mu}.
\begin{lem}\label{lem:mu2}
For $0< b<\iny$,
\ben\lel{eq:lem_mu2}
\sup_{\theta\in \cN_{b}(\theta_{0})}\stj\stk\left[ \sti d_{ij}\left\{ f_{i}(k;\bth)-f_{i}(k;\bth_{0}) -  (\bth-\bth_{0})'\bqi(k,\bth_{0}) \right\} \right]^2 = o(1),
\een
where $\bqi(k,\bth):=\partial f_{i}(k;\bth)/\partial \bth\in \mR^{p}$.
\end{lem}
\begin{proof}
To begin with, note that $\bqi(k,\bth)=\bX_{i}\bg_{i}(k,\bth)$, which is an analogue of $g_{i}\bxi$ of Section 2. Replacing it with one in (\ref{eq:uAX}), application of the mean value theorem (MVT) yields that
\benn
f_{i}(k;\bth)-f_{i}(k;\bth_{0})=
\bu_{1}'\mbf{c}_{1i}g_{1i}(k;\wt{\bth}) +  \bu_{2}'\mbf{c}_{2i}g_{2i}(k;\wt{\bth}),
\eenn
where $\wt{\bth}=(\wt{\B}',\wt{\D}')'=c\bth+(1-c)\bth$ for some $0\le c\le 1$. Thus
\benrr
f_{i}(k;\bth)-f_{i}(k;\bth_{0}) -  (\bth-\bth_{0})'\bqi(k,\bth_{0})
&=& \bu_{1}'\mbf{c}_{1i}\big[g_{1i}(k,\wt{\bth})-g_{1i}(k,\bth_{0}) +\bu_{2}'\mbf{c}_{2i}\big[g_{2i}(k,\wt{\bth})-g_{2i}(k,\bth_{0})\big].  \eenrr
With one more application of of the MVT implies that
\benn
g_{1i}(k,\wt{\bth})-g_{1i}(k,\bth_{0})= \bxii'(\wt{\B}-\B_{0})\la_{i}\frac{\partial g_{1i}(k,\bth_{*})}{\partial \la_{i}^2}+\bxiii'(\wt{\D}-\D_{0})\frac{\partial g_{1i}(k,\bth_{*})}{\partial \vp_{i}},
\eenn
where $\bth_{*}$ is such that $\|\bth_{*}-\bth_{0}\|\le \|\wt{\bth}-\bth_{0}\|$, and hence, we have
\benrr
\max_{1\le i\le n} \stk|g_{1i}(k,\wt{\bth})-g_{1i}(k,\bth_{0})|^2&=&\max_{1\le i\le n} \stk\left|\bxii'(\wt{\B}-\B_{0})\la_{i}\frac{\partial g_{1i}(k,\bth_{*})}{\partial \la_{i}^2}+\bxiii'(\wt{\D}-\D_{0})\frac{\partial g_{1i}(k,\bth_{*})}{\partial \vp_{i}}\right|^2,\\
&\le & 4c^2 b^2 \max_{1\le i\le n} \|\mbf{c}_{1i}\|^2\left[\la_{i}^{2}\stk\left(\frac{\partial f_{i}(k;\bth_{*})}{\partial \la_{i}}\right)^2+\la_{i}^4 \stk\left(\frac{\partial^2 f_{i}(k;\bth_{*})}{\partial \la_{i}^2}\right)^2\right]\\
&&\quad + 4c^2 b^2\max_{1\le i\le n}\|\mbf{c}_{2i}\|^2\la_{i}^2 \stk\left(\frac{\partial^2 f_{i}(k;\bth_{*})}{\partial \la_{i}\partial \vp_{i}}\right)^2\ra 0,
\eenrr
where the inequality follows from $\wt{\bth}=c\bth+(1-c)\bth_{0}$, $\|\bu_{l}\|\le b,\,l=1,2$, and $(a+b)^2\le 2(a^2+b^2)$ for $a,b\in \mR$ while the convergence to 0 follows from (\tbf{a.3})', (\tbf{a.5})', and (\ref{eq:gk_bdd3}). Using the same argument, the convergence of $\max_{1\le i\le n} \stk|g_{1i}(k,\wt{\bth})-g_{1i}(k,\bth_{0})|^2$ to zero can be shown. Finally, we have
\benrr
&&\textrm{the supremand in (\ref{eq:lem_mu2})}\\
&\le&  \stj \stk\left[ \sti |d_{ij}\mbf{u}_{1}'\mbf{c}_{1i}|\cdot |g_{1i}(k;\wt{\bth})-g_{1i}(k;\bth_{0})|
+\sti |d_{ij}\mbf{u}_{2}'\mbf{c}_{2i}|\cdot |g_{2i}(k;\wt{\bth})-g_{2i}(k;\bth_{0})|
\right]^{2},\\
&\le&  2\stj b^2 \left[ \sti \|d_{ij}\mbf{c}_{1i}\|^2\cdot
\max_{1\le i\le n}\stk|g_{1i}(k;\wt{\bth})-g_{1i}(k;\bth_{0})|^2 \right. \\
& & \qquad \qquad \left.
+\sti \|d_{ij}\mbf{c}_{2i}\|^2\cdot
\max_{1\le i\le n}\stk|g_{2i}(k;\wt{\bth})-g_{2i}(k;\bth_{0})|^2 \right]\ra 0,
\eenrr
where the first inequality follows from the triangle inequality while $\|\bu_{l}\|\le b,\,l=1,2$ and $(a+b)^2\le 2(a^2+b^2)$ for $a,b\in \mR$ again imply the second inequality. Then, (\tbf{a.4})' will yield the convergence to zero, thereby completing the proof of the lemma.
\end{proof}

\section{Conclusion}
In this study, we extended the application of the CvM type distance -- which is popular in the continuous probability distributions -- to a Poisson one sample and regression setups and proposed the MD estimators through using its analogue, that is, with the integral of the original CvM type distance being replaced by the summation. Based on the promising results shown in this article, further extension to broad range of discrete probability distributions and and to other statistical model is expected to yield some desirable results, and hence, will form future research.

\bibliographystyle{plainnat}
\bibliography{MDE_ref.bib}

\edt